\documentclass[9pt,twocolumn,twoside]{opticajnl}
\journal{opticajournal} 

\setboolean{shortarticle}{true}
\usepackage{subcaption}
\usepackage{lineno}

\title{Annular beams for reliable intersatellite optical communications}

\author[1,*]{Mario Badás Aldecocea}
\author[1]{Edward Pauwels}
\author[1]{Jasper Bouwmeester}
\author[1]{Pierre Piron}
\author[1]{Jérôme Loicq}

\affil[1]{Department of Space Engineering, Delft University of Technology, 2629 HS Delft, the Netherlands}
\affil[*]{mbadasaldecoce@tudelft.nl}

\begin{abstract}
Free-space optical communications (FSOC) are a key enabling technology for future high-capacity space-based networks. Particularly, the backbone of global communication relies on intersatellite optical links.
In a previous study \cite{badas_optimum_2024}, the authors proposed a method to mitigate the impact of transmitter pointing jitter by using a superposition of orthogonally polarized Gaussian and higher-order Laguerre-Gaussian (LG) beams.
In this study, we experimentally characterize the proposed system using a spiral phase plate (SPP) to generate higher-order annular beams.
We demonstrate that such superpositions can be reliably generated in a realistic optical setup, quantify the associated beam-shaping errors and losses, and assess their impact on intersatellite optical communication performance. It is found that the proposed beam-shaping approach can still yield power savings on the order of 20\% compared to a conventional Gaussian beam under the considered conditions.
\end{abstract}

\setboolean{displaycopyright}{false} 

\begin{document}

\maketitle

\section{Introduction}

To enable a global, high-speed, and reliable communication infrastructure, intersatellite free-space optical communication links are of paramount importance. Compared to their radio-frequency counterparts, FSOC terminals offer significantly higher data throughput while requiring much smaller antenna sizes. However, FSOC satellite transmitters are affected by pointing jitter arising from several space-environmental and internal effects \cite{badas_opto-thermo-mechanical_2023}, such as reaction wheel microvibrations or micrometeorite impacts. To minimize the associated losses, active pointing and tracking systems are employed in these terminals. Nevertheless, such mechanisms are inherently imperfect, and the communication link is always subject to a residual pointing jitter.

This residual jitter induces power fluctuations at the receiver, which degrade the overall performance of the communication link. A first mitigation approach, proposed by Toyoshima \textit{et al.} \cite{toyoshima_optimum_2002}, consists of optimizing the divergence of a transmitted Gaussian beam. Extending this idea, we previously proposed to consider beam shapes beyond the fundamental Gaussian mode \cite{badas_optimum_2024}. Specifically, we introduced the use of a superposition of orthogonally polarized Gaussian and higher-order Laguerre-Gaussian beams to improve system performance. As will be presented in this paper, the generated annular beams --created by impinging a Gaussian beam onto a spiral phase plate-- do not exactly correspond to LG beams. However, the performance gain provided by these beams, relies on the same principle. The performance gain relies on redistributing the total transmitted power over a broader spatial profile, thereby reducing the sensitivity of the received signal to pointing-induced power fluctuations. In this way, the system deliberately sacrifices peak received power in favor of a more stable and reliable signal, which ultimately improves link-level performance metrics such as outage probability and average bit error probability under pointing jitter. Under the assumption of ideal beam shapes, this approach was shown to yield power savings on the order of $20$ to $40\;\%$ .

In this letter, we present the first experimental demonstration of the proposed concept. The aim of this work is to characterize the practical implementation of the system, with particular emphasis on the beam-shaping performance of the SPP used to generate annular beams and on the power losses introduced by a realistic optical setup. This paper is structured as follows. Section~\ref{sec_setup} describes the experimental setup and the beam-shaping principle. Section~\ref{sec_results} presents the experimental results, including beam shape quality and power losses. Section~\ref{sec_performance} evaluates the impact of the experimentally obtained beams on the performance of an intersatellite FSOC link.

\section{Experimental setup}\label{sec_setup}
\begin{figure}[!htb]
    \centering
    \includegraphics[width=\linewidth]{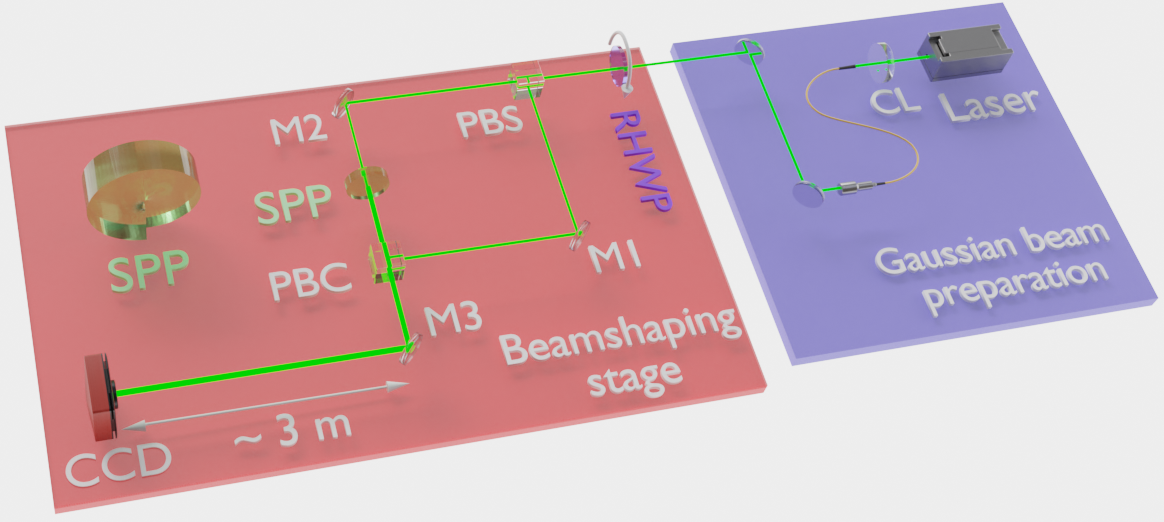}
    \caption{Experimental setup proposed for beam-shaping. The Gaussian beam cleaning stage and the beam-shaping stage are shown. A picture of the experimental setup in the laboratory can be found in the Supplemental Document.}
    \label{fig_setup}
\end{figure}

To generate a superposition of orthogonally polarized Gaussian and higher-order annular beams, the experimental setup shown in Fig.~\ref{fig_setup} is employed at a wavelength of $532\;\mathrm{nm}$. Minor modifications with respect to the setup proposed in Ref.~\cite{badas_optimum_2024} were implemented, the most important being the addition of a wavefront-cleaning stage based on a single-mode fiber. This stage ensures a high-quality Gaussian beam at the input of the beam-shaping section. As shown in the Supplemental Document, the SPP used to generate the annular beam is highly sensitive to the quality of the incident Gaussian beam. In a realistic FSOC terminal, such an additional wavefront-cleaning stage would typically not be required, since most FSOC sources are already fiber-coupled, for instance due to the use of fiber-based modulators and erbium-doped fiber amplifiers \cite{carrasco-casado_development_2022, plooy_cubecat_2025, ruddenklau_-orbit_2024}.

After the wavefront-cleaning stage, the Gaussian beam from the fiber collimator passes through a periscopic pair of mirrors. The beam is linearly polarized, and its polarization orientation is adjusted using a rotatable half-wave plate (RHWP). By changing the angle between the incoming polarization and the fast axis of the RHWP, the polarization state incident on the polarization beam splitter (PBS) is controlled. This, in turn, determines how the optical power is split between the two arms of the setup. The horizontally polarized Gaussian beam --transmitted by the PBS-- then passes through the SPP. The fused silica SPP is a diffractive optical element that imposes a helicoidal phase profile on the incoming Gaussian beam. After sufficient propagation distance, this phase modulation leads to the formation of an annular beam shape \cite{ruffato_generation_2014, massari_fabrication_2015}.

Although other techniques exist to generate annular beam shapes, such as spatial light modulators or birefringent vortex plates \cite{yao_orbital_2011}, these approaches often introduce additional complexity, dynamic behavior, or higher losses, which are undesirable for FSOC applications. The SPPs used in this work generate annular beams of topological orders $\ell = 1, 2, 3$, which are the most relevant modes for pointing-jitter mitigation according to Ref.~\cite{badas_optimum_2024}.

The vertical polarization component reflected by the PBS bypasses the SPP and is subsequently superposed with the horizontally polarized annular beam using a polarization beam combiner (PBC). The resulting beam is then propagated over a distance of more than $\sim3$~m to allow the annular beam shape to fully develop. The resulting irradiance pattern is measured using a camera. Since the Gaussian and annular components are orthogonally polarized, the measured irradiance corresponds to the incoherent sum of their individual irradiance profiles. By rotating the RHWP, the power ratio between the Gaussian and annular components can be continuously adjusted, thereby generating a family of composite beam shapes. A video illustrating this evolution is provided in {Visualization~1}.

To assess the quality of the experimentally generated beams, the measured irradiance patterns are compared with numerical simulations. These simulations are based on a Fourier-optics propagation model capable of describing both near- and far-field propagation \cite{goodman_introduction_2005, schmidt_numerical_2010}, combined with Jones calculus to model polarization effects throughout the setup \cite{collett_field_2005}. Further details of the developed model are provided in the Supplemental Document.

\section{Results}\label{sec_results}

In this section, we present and analyze the experimentally obtained beam shapes and quantify the power losses introduced by the proposed optical system. The experimental results are systematically compared with numerical simulations in order to assess the beam-shaping performance of the setup.

For a meaningful comparison between simulations and experiments, the Gaussian beam emitted by the fiber collimator was first characterized. This was accomplished by matching simulated results to experimental beam profiles for different assumed Gaussian beam waists and wavefront curvatures at the collimator output. From this procedure, the beam was found to have a waist of $w_{0}=1.19\;\text{mm}$ and a wavefront radius of curvature of $R_{\text{col}}=12\;\text{m}$. Details of the retrieval procedure are provided in the Supplemental Document.

\subsection{Beam-shaping}

\begin{figure}[h!]
    \centering
    \includegraphics[width=\linewidth]{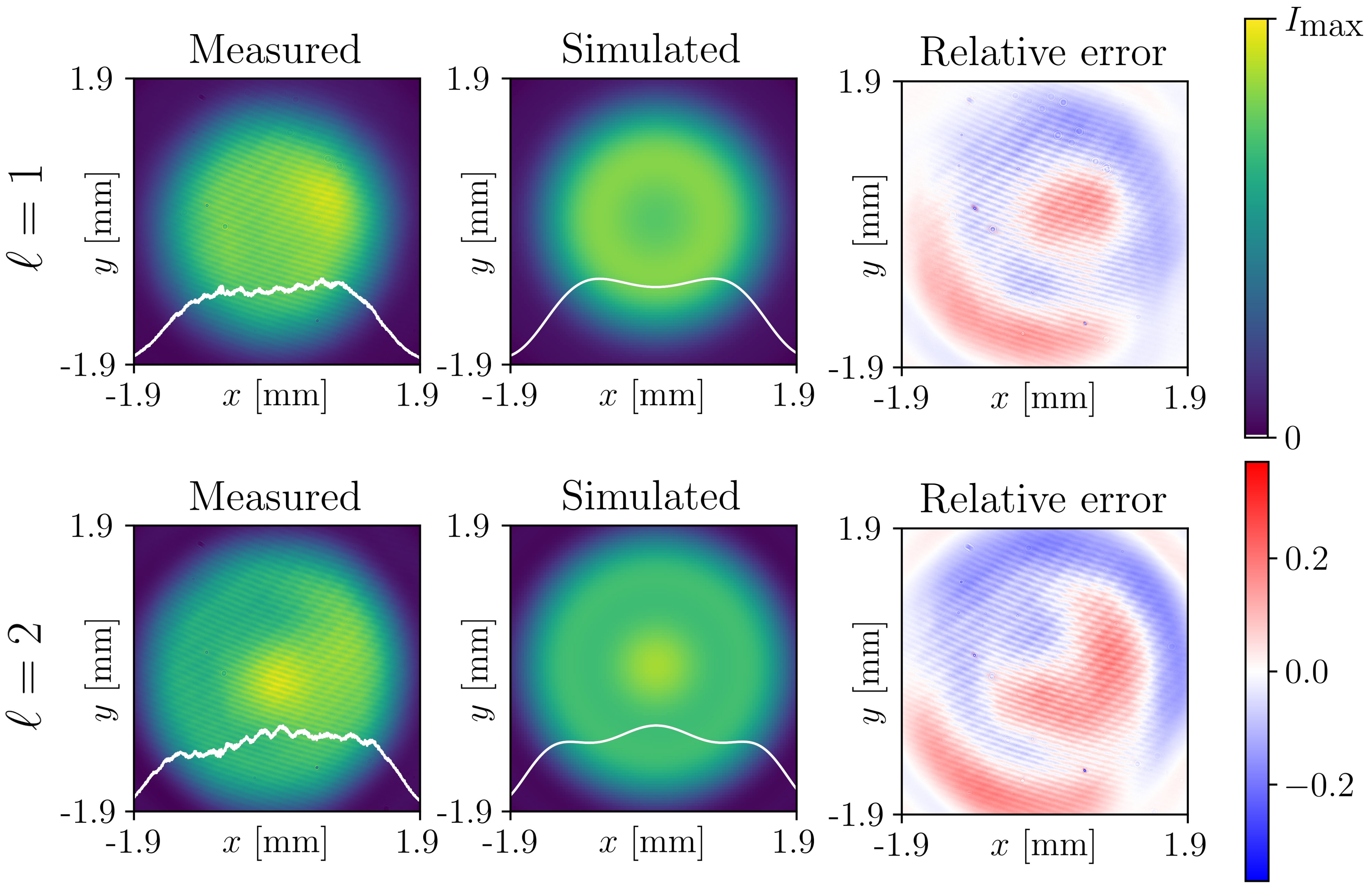}
    \caption{Measured and simulated beam shapes for different topological orders, with equal power contributions (RHWP angle $22.5^\circ$) from each beam (Gaussian and annular). The white lines indicate the irradiance cross-section at $y=0$.}
    \label{fig_result_beamshape}
\end{figure}

After characterizing the Gaussian beam at the aperture of the fiber collimator, simulated and measured beam shapes can be directly compared. Both the simulated and experimental beams are normalized to unit power, enabling a comparison focused solely on beam shape. Figure~\ref{fig_result_beamshape} shows the results for superpositions of Gaussian and annular beams with topological orders $\ell=1$ and $2$ when the RHWP is set to $22.5^\circ$, corresponding to equal power splitting between the two beams (an extended set of figures for other RHWP angles and $\ell$ can be found in the Supplemental Document). The residuals between the simulated and experimental irradiance patterns are also shown. These residuals reveal a slight asymmetry in the experimental annular beams, with increased power in the top-right region compared to the bottom-left. As will be shown later in Section~\ref{sec_performance}, these small asymmetries do not lead to significant losses in the communication performance of the generated beam shapes.

\begin{figure}
    \centering
    \includegraphics[width=\linewidth]{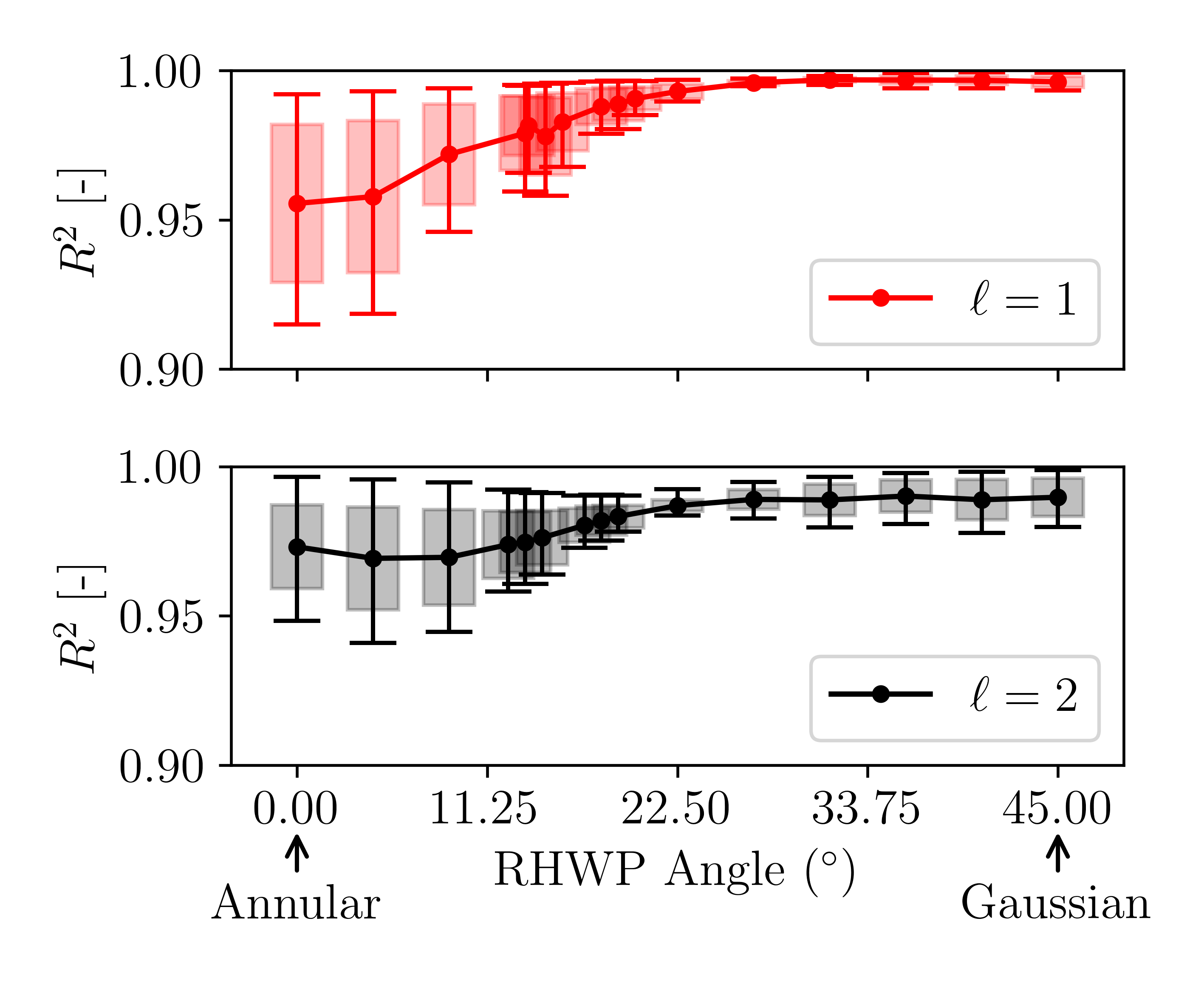}
    \caption{$R^2$ values obtained by comparing simulated and experimental beam shapes for different RHWP angles, for $\ell=1$ and $\ell=2$ annular beams. Average (dots), standard deviation (boxes) and extremal values (whiskers) are shown.}
    \label{fig_results_r2}
\end{figure}

As discussed in Ref.~\cite{badas_optimum_2024}, the optimal power distribution between the Gaussian and annular components depends on the chosen performance metric and the operating conditions (e.g., pointing jitter, available power, etc.). Therefore, it is important to characterize the beam-shaping performance across the full range of RHWP angles. This is done by computing an $R^2$ metric that quantifies the agreement between simulated and experimental irradiance patterns. Specifically, for each RHWP angle, the slice-wise coefficient $R^2_\mathrm{slice}=1-\iint|I_\mathrm{sim}-I_\mathrm{exp}|^2\,dx\,dy$ is computed over multiple angular slices of the beam. Figure~\ref{fig_results_r2} shows the average $R^2$ across all slices, along with the standard deviation and the extremal values, as a function of RHWP angle for $\ell=1$ and $\ell=2$.

Overall, the average $R^2$ remains above 95\%, indicating that the intended beam shapes are reproduced with high fidelity, being this higher the higher the contribution of the Gaussian beam to the final shape. The observed discrepancies between simulated and experimental beam shapes can be attributed to several factors. Although the wavefront-cleaning stage significantly improves the Gaussian beam quality (see Suplemental Document), residual wavefront aberrations remain and are transferred to the annular beam generated by the SPP. In addition, the beam quality is sensitive to manufacturing variations between different SPP batches. Finally, spurious interference fringes caused by internal reflections within the camera system introduce additional distortions in the measured irradiance patterns.

The results presented are valid on the camera plane of the proposed setup ($\sim 3$~m away from the SPP, as shown in Fig.~\ref{fig_setup}). The further development of the annular beam after the SPP is discussed in detail in the Supplemental Document. Both simulations and measurements show that the beam continues to evolve with propagation distance, approaching an annular profile that differs from a Laguerre–Gaussian beam. The analysis in the Supplemental Document demonstrates that the SPP-generated beam does not exactly correspond to a Laguerre–Gaussian mode, with a larger transverse size and a distinct far-field profile; achieving an exact LG mode would require an additional transmission screen, which would introduce significant power losses. The subsequent alignment sensitivity analysis shows that the beam is highly tolerant to angular misalignments of SPP but extremely sensitive to lateral displacements at the micrometer scale. This indicates that precise lateral positioning of the SPP is critical for maintaining the desired beam shape (see Supplemental Document for details).

\subsection{Power losses}

In addition to beam-shaping performance, it is essential to quantify the optical losses introduced by the proposed setup, as these directly impact the achievable communication performance. Here, we focus only on losses in the beam-shaping stage shown in Fig.~\ref{fig_setup}, since the Gaussian beam cleaning stage is not representative of a realistic FSOC terminal, where fiber coupling is typically already present due to the use of fiber-based modulators and amplifiers \cite{carrasco-casado_development_2022, plooy_cubecat_2025, ruddenklau_-orbit_2024}.

The dominant sources of loss arise from the SPP. These include Fresnel reflections, absorption, and scattering due to fabrication imperfections. Other optical components, such as the RHWP and polarization beam splitters, contribute polarization-dependent insertion losses; however, these are negligible compared to the losses introduced by the SPP. To quantify the SPP-induced losses, power measurements were performed with a power meter. The measured transmission efficiencies are $92.6\%$ for $\ell=1$ and $87.5\%$ for $\ell=2$. The annular beam with $\ell=3$ was generated experimentally by cascading SPPs with $\ell=1$ and $\ell=2$, which leads to an accumulated transmission loss of $81\%$. In a practical implementation, the latter could be significantly reduced by using a single SPP with topological order $\ell=3$, which would avoid cascading elements and reduce the transmission loss. In the following section, we analyze the experimentally obtained beam shapes both with and without accounting for these measured power losses, evaluating their expected performance in an intersatellite optical communication link.

\section{Intersatellite optical communication}\label{sec_performance}

\begin{figure}
    \centering
    \includegraphics[width=0.95\linewidth]{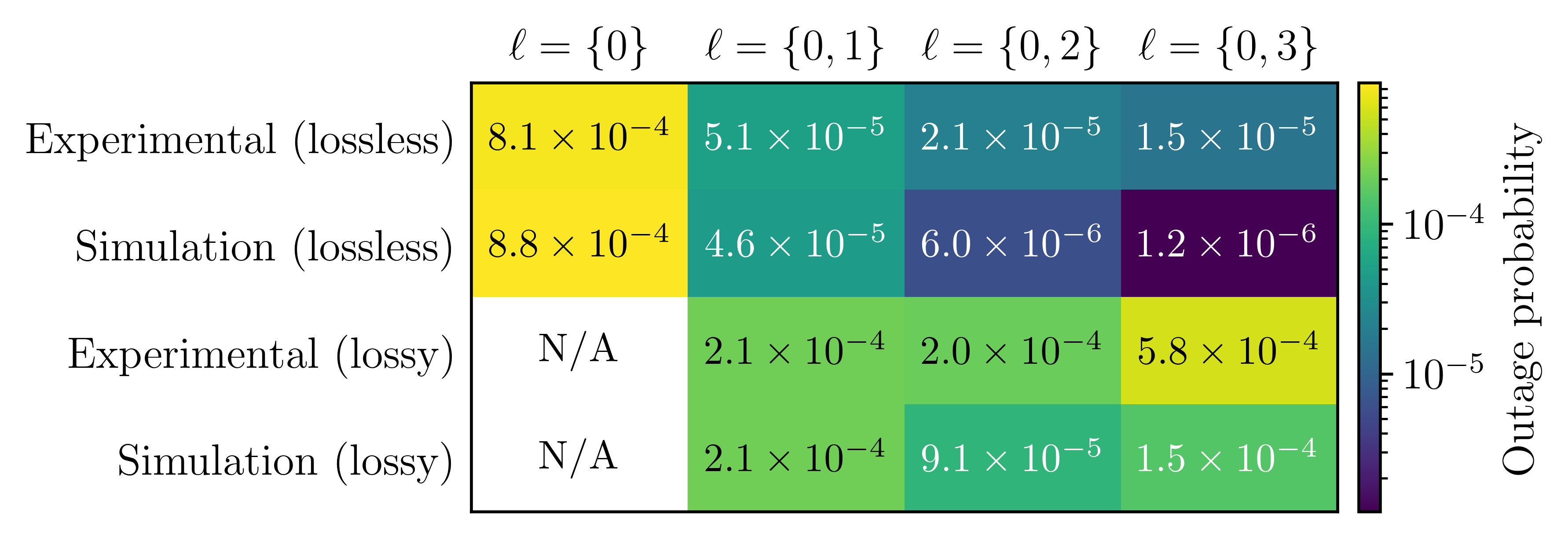}
    \caption{Outage probability performance for the measured and simulated beam shapes at the camera plane. The beam shapes are clipped by the camera size and are not fully developed yet.}
    \label{fig_communication_performance_pout_camera}
\end{figure}

Using the experimentally obtained beam profiles, the performance of an intersatellite FSOC link is evaluated through numerical simulations. Two scenarios are considered. In the first, the power losses discussed in the previous section are neglected, which allows the isolated effect of beam-shaping on pointing-jitter mitigation to be assessed. In the second, the measured losses are included, providing a realistic estimate of the current system performance. The model used to evaluate the communication performance is described in Section 2 of Ref.~\cite{badas_optimum_2024}. The intersatellite link parameters employed in the simulations are taken from Table~1 of the same reference. For each beam configuration, the optimum performance is obtained by simulating different beam sizes and different relative contributions of the Gaussian and annular components, and then identifying the configuration that minimizes the outage probability. As explained in Ref.~\cite{badas_optimum_2024}, the outage probability is defined as the probability that the instantaneous received power falls below the threshold required for an ideal receiver to sustain communication at a prescribed data rate. The results obtained are summarized in Fig.~\ref{fig_communication_performance_pout_camera}, which shows the outage probability for Gaussian beams and for superpositions of Gaussian and annular beams of different topological orders at the camera plane. It is important to note that these beams do not correspond to those encountered at the receiver aperture, since they are affected by the finite clipping of the camera and are not yet fully developed. Nevertheless, this comparison is useful to further validate the simulation framework by allowing the measured fields to be propagated both numerically and analytically to the far field and to assess the resulting communication performance. These results already highlight the benefits of using superpositions of Gaussian and annular beams compared with a conventional Gaussian beam. In particular, the outage probability is significantly reduced, even when the power losses induced by the SPP are taken into account. This demonstrates the potential of tailored beam-shaping as an effective means of mitigating pointing-jitter impairments in practical intersatellite FSOC links.

\begin{figure}
    \centering
    \includegraphics[width=\linewidth]{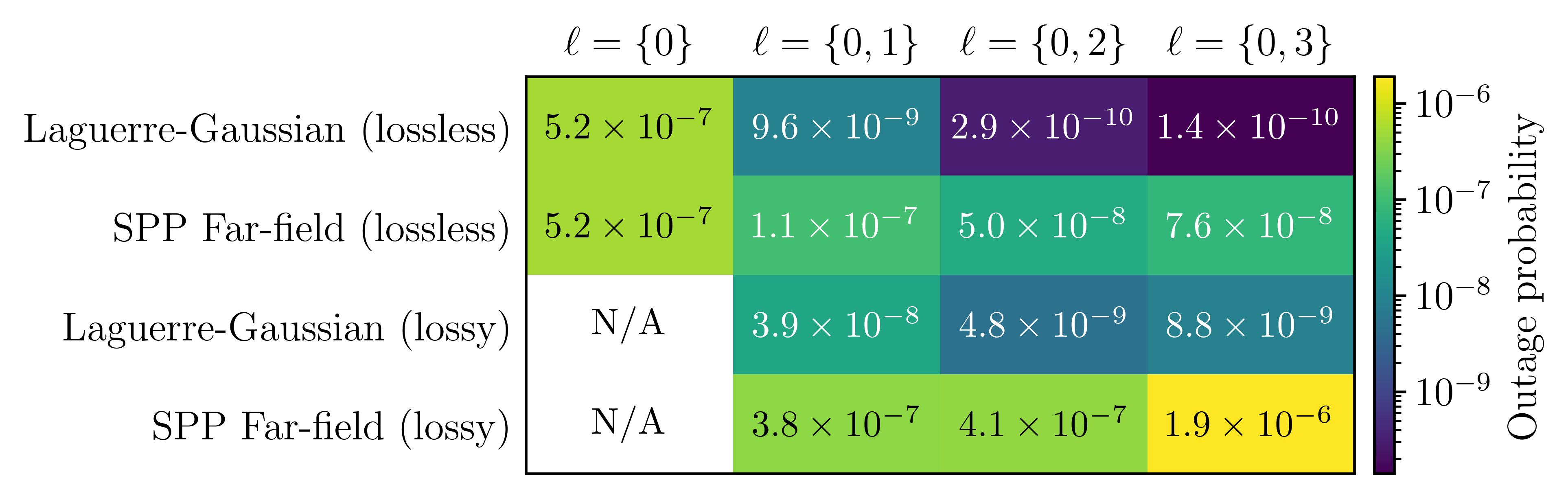}
    \caption{Outage probability performance for the analytical and simulated beam shapes at the far-field. The analytical and numerical far-field shapes obtained by the SPP, along with the optimal shapes obtained if LG beams were generated, are shown.}
    \label{fig_communication_performance_pout_ff}
\end{figure}

As can be seen in Fig.~\ref{fig_communication_performance_pout_camera}, the good agreement between the measured and simulated beam profiles shown in Fig.~\ref{fig_results_r2} is also successfully reflected in the predicted communication performance. This consistency provides confidence in the simulation framework and enables the results to be propagated further to the far field, where the beam shapes expected at the receiver aperture plane can be used to evaluate the intersatellite link performance under realistic conditions. The resulting outage probabilities for the far-field–simulated beam shapes are presented in Fig.~\ref{fig_communication_performance_pout_ff}. In this case, the superpositions of LG far fields (see Ref.~\cite{badas_optimum_2024}) are compared with analytical far-field expressions obtained for an SPP-generated annular beam superposed with a Gaussian beam. The analytical far-field irradiance of an SPP-generated annular beam as a function of the radial coordinate $\rho$ is given by (see Supplemental Document for the derivation)
\begin{equation*}
    I_\text{FF}(\rho,z)\propto  e^{-\frac{\rho^2}{w^2(z)}} \left| \frac{ \rho }{z^2 } \left[\mathcal{I}_{\frac{\ell
   -1}{2}}\left(\frac{\rho^2}{2 w^2(z)}\right)-\mathcal{I}_{\frac{\ell +1}{2}}\left(\frac{\rho^2}{2
   w^2(z)}\right)\right]\right|^2
\end{equation*}
where $w(z)$ is the Gaussian beam radius at propagation distance $z$ (considering the geometry of the Gaussian beam impinging at the SPP at the beamwaist), and $\mathcal{I}_n(x)$ denotes the modified Bessel function of the first kind of order $n$. The analytical far-field beam shapes corresponding to the experimentally generated fields are then obtained by a weighted superposition --set by the RHWP angle-- of the Gaussian beam irradiance and the annular beam irradiance given by the above expression. These analtically obtained beams match the profiles obtained by wave propagation simulations as can be seen in the Supplemental Document. This approach allows to translate the experimental-simulation match found in Fig.~\ref{fig_communication_performance_pout_camera} into the far-field analytical irradiance profiles, and enables a reliable assessment of the communication performance using the beam profiles expected at the receiver aperture. The results in Fig.~\ref{fig_communication_performance_pout_ff} demonstrate the improvement, relative to a conventional Gaussian beam, provided by the beam-shaping technique proposed in this chapter for mitigating the effects of transmitter pointing jitter in intersatellite optical communication. Furthermore, the table shows that although the SPP-generated annular beams do not perform as well as LG beams, they still provide a significant enhancement in system performance compared to the conventional Gaussian beam. Notably, even the power losses introduced by absorption in the SPP are compensated by the proposed beam-shaping approach, except in the $\ell=3$ case, where the additional losses arise from the use of two cascaded SPPs. In terms of power efficiency, these results indicate that power savings of up to 20\% could be achieved if the SPP losses were further minimized, for example through the use of anti-reflective coatings or other loss-reduction techniques. The reader is referred to the Supplemental Document for obtaining the performance of these beams in terms of the average bit error probability.


Future work should focus on reducing system losses through improved antireflection coatings on the SPP and on miniaturizing the beam-shaping setup for integration into space-qualified FSOC terminals. An end-to-end experimental validation using a communication terminal with induced pointing jitter is also envisaged. Beyond the present concept, alternative beam-shaping techniques and other structured beam profiles should be investigated for further performance improvements.



\begin{backmatter}
\bmsection{Funding} 


\bmsection{Disclosures} The authors declare no conflicts of interest.

\bmsection{Data availability} Data underlying the results presented in this paper are not publicly available at this time but may be obtained from the authors upon reasonable request.

\bmsection{Supplemental document}
See Supplemental Document for supporting content.
See Visualization 1 for the video of the combination of the Gaussian and annular beam superposition as the RHWP rotates.
\end{backmatter}

\bibliography{references}

\bibliographyfullrefs{references}

\end{document}